\documentclass[twocolumn,aps,amsfonts,prl]{revtex4}
\usepackage{graphicx}
\usepackage{mathrsfs}
\usepackage{amsmath}
\usepackage{mathrsfs}
\usepackage{amssymb}
\usepackage{color}
\usepackage[nice]{nicefrac}
\usepackage{float}
\usepackage{times}
\begin{document}

\def\be{\begin{eqnarray}}
\def\ee{\end{eqnarray}}
\def\nn{\nonumber}
\newcommand{\note}[1]{$\color{red}{\bullet}$}
\def\mpp#1{{\marginpar{\tiny #1}}}

\title{Fractional Quantum Hall Effect of Lattice Bosons Near Commensurate Flux}

\author{L. Hormozi$^1$}
\author{G. M\"oller$^2$}
\author{S. H. Simon$^3$}
\affiliation{
$^1$Joint Quantum Institute, National Institute of Standards and Technology and University of Maryland, Gaithersburg, Maryland 20899, USA\\
$^2$TCM Group, Cavendish Laboratory, University of Cambridge, Cambridge CB3 0HE, United Kingdom \\
$^3$Rudolf Peierls Centre for Theoretical Physics, University of Oxford, Oxford OX1 3NP, United Kingdom }

\begin{abstract}
We study interacting bosons on a lattice in a magnetic field.  When the number of flux quanta per plaquette is close to a rational fraction, the low-energy physics is mapped to a multispecies continuum model: bosons in the lowest Landau level where each boson is given an internal degree of freedom, or \emph{pseudospin}.
We find that the interaction potential between the bosons involves terms that do not conserve pseudospin, corresponding to umklapp processes, which in some cases
can also be seen as BCS-type pairing terms.  We argue that in experimentally realistic regimes for bosonic atoms in optical lattices with synthetic magnetic fields, these terms are crucial for determining the nature of allowed ground states. In particular, we show numerically that certain paired wave functions related to the Moore-Read Pfaffian state are stabilized by these terms, whereas certain other wave functions can be destabilized when umklapp processes become strong.
\end{abstract}

\maketitle

Recent advances in the field of topological phases and their potential application in implementing an intrinsically fault-tolerant quantum computer~\cite{kitaev03,nayak08} have revitalized interest in fractional quantum Hall (FQH) states as the most prominent examples of topologically ordered phases of matter~\cite{wen90}.  Even though it has only been observed in (fermionic) solid-statesystems, the FQH effect can also exist for bosons~\cite{wilkin00, cooper01, cooper08, morris07}.  Promising candidates are systems of interacting ultra-cold atoms where the necessary magnetic fields are simulated by   rapid rotation~\cite{madison00,abo01} or by laser-induced synthetic gauge fields~\cite{laserinduced}.  At low temperatures when the filling fraction $\nu$ (the ratio of the particle density $n$ to the magnetic flux density $n_\phi$) is sufficiently small, one can expect to observe bosonic versions of the FQH effect~\cite{cooper01}. For example, the exact ground state of bosons with contact interaction at filling fraction $\nu = 1/2$ is the Laughlin state~\cite{wilkin00, paredes01}, while at $\nu = 1$ the ground state is in the same topological phase as the non-Abelian Moore-Read Pfaffian state ~\cite{mooreread, cooper01}.

A major advantage of optical and atomic systems over conventional solid-state systems is the possibility of creating and controlling quasiparticle excitations more naturally and with higher precision (e.g., by shining focused laser beams on the atomic gas)~\cite{paredes01}. A number of proposals suggest that the FQH regime for cold atoms can be most easily achieved using optical lattices~\cite{LatticeFlux1,LatticeFlux2,sorensen05, gerbier10, cooper11}.  The question naturally occurs whether there is new physics that may arise for a system of interacting bosons in the FQH regime due to the effects of an underlying lattice. It has been shown that in the limit when the flux density $n_\phi$, or equivalently the number of flux quanta per lattice plaquette, is small, one can ignore the existence of the lattice and treat the system in the continuum limit~\cite{sorensen05, hafezi07}. When $n_\phi$ is large, however, the presence of the lattice can potentially lead to new correlated states of matter that are absent in the continuum~\cite{palmer06, palmer08, moller09}. This is the limit we will focus on.

The starting point for our analysis of the many-body physics in this problem is the observation that when $n_\phi$ is close to a rational fraction, the lowest energy bands in the Hofstadter butterfly, a fractal structure realizing the single-particle energy spectrum of particles hopping on a lattice in a magnetic field~\cite{hofstadter76},  are reminiscent of Landau levels in the continuum. This resemblance can be formalized by mapping the single-particle states of the system to a continuum model when the flux density is near simple rational fractions~\cite{palmer06}.

The main result of this Letter is the following.  In agreement with~\cite{palmer06}, we find that for flux per plaquette close to a rational fraction, $n_\phi=p/q + \epsilon$ with $p,q$ small integers, and $\epsilon$ sufficiently small, one can map the system to an effective continuum model with Landau levels and an added degree of freedom for the particles, a sub-band index or pseudospin, which can take $q$ possible values. However, in addition to the density-density interactions between bosons of different pseudospin found in~\cite{palmer06}, we find anomalous ``pairing" interactions that do not conserve the number of particles of each pseudospin species.  We find that these  pairing terms, which become increasingly strong as $\epsilon$ is increased, are crucial in determining the possible  ground states of the system for realistic values of the parameters of the problem.

As a detailed example, we consider the most (experimentally) realistic case $n_\phi = 1/2 + \epsilon$ and study several effective filling fractions $\tilde \nu = n/\epsilon$. We find a new FQH state at $\tilde \nu = 1$,  which does not exist without the  pairing interactions but becomes stabilized by the increase in $\epsilon$ and the concomitant increase in these interactions. This new state is related to two copies of the non-Abelian Moore-Read Pfaffian state~\cite{mooreread}; however, it is a topologically distinct phase of matter.  In contrast, we find that the pairing terms {\it destabilize} the states predicted at fillings $\tilde \nu=2/3$ previously discussed by~\cite{palmer06}, $\tilde \nu = 4/3$~\cite{palmer08}, and $\tilde \nu=2$~\cite{moller09}.   We present detailed numerical evidence for our conclusions and argue that experiments are most likely to be in a regime where these  pairing terms are important.

We consider bosons with onsite repulsive interaction, hopping on a two-dimensional square lattice, subject to a uniform perpendicular effective magnetic field. This system is described by a modified Bose-Hubbard Hamiltonian~\cite{jaksch98},
\be
H = -J\sum_{<ij>} \left(e^{i\theta_{ij}}c^\dagger_i c_j +  h.c.\right) + U\sum_{i}  c^\dagger_i c^\dagger_i c^{\phantom{\dagger}}_i c^{\phantom{\dagger}}_i\label{ham},
\ee
where $c_{i}^{\dagger}$  and $c_i$ are boson creation and annihilation operators on lattice site $i$, $J$ is the hopping energy, and $U$ is the strength of the onsite interaction.  Here $\theta_{ij} = \int_i^j\vec{A}\cdot\mathrm{d}\vec{l}$ is the phase acquired by a particle hopping from site $i$ to the neighboring site $j$ with $\vec{A}$ being the vector potential, and we work in units where $\hbar = 1$, and the effective electric charge coupled to $\vec{A}$ is also set to unity. The kinetic term in the Hamiltonian then indicates that a particle hopping around a lattice plaquette acquires a phase of $2\pi n_\phi$.

We start by considering the kinetic term of the Hamiltonian only. This is the well-known   single-particle Hofstadter problem, which we review only briefly.  We assume the lattice is in the $x,y$-plane, with lattice spacing set to unity for simplicity, and choose the Landau gauge so that $\vec{A} = (0,2 \pi n_{\phi}\, x,0)$. The wave function becomes $\psi(x,y) = \phi(x)e^{iky}$, where $\phi(x)$ satisfies Harper's equation $\phi(x+1) + \phi(x-1) + 2 \cos(2 \pi n_\phi x - k) \phi(x) =  (E/J) \phi(x)$ and $k$ is the momentum in the $y$-direction. Note that $x$ and $y$ are both integers.

Consider the case of $n_\phi = \epsilon \ll 1$, where a continuum approximation of the discrete Harper's equation can be used for the low-energy eigenstates.    In this limit, it is convenient to use a Wannier basis localized near minima of the cosine potential. These Wannier functions can be approximated by harmonic oscillator (Landau level) solutions with oscillator length (magnetic length) $l_0 = 1/\sqrt{2\pi\epsilon}$ centered at $x_k = k/(2\pi\epsilon)$,  i.e., $\phi_k(x) \sim \exp(-\pi\epsilon(x - x_k)^2)$.  The bandwidth of the lowest band arises from tunneling between adjacent minima of the potential and for small $\epsilon$ it scales as $\sim e^{-C/\epsilon}$ where the constant $C\approx 1.166$ can be obtained by the WKB approximation~\cite{Watson91}.  Note that for small $\epsilon$, the bandwidth is much smaller than the band gap $\Delta= 4J\pi\epsilon$, making this limit of the Hofstadter problem an example of an (almost) flat Chern band~\cite{FlatBand}.

Now let us consider flux densities close to a rational fraction. For simplicity, we focus on $n_\phi = 1/2 + \epsilon$, for which Harper's equation becomes $\phi(x+1) + \phi(x-1) + 2 (-1)^{x} \cos(2\pi \epsilon x - k) \phi(x) =  (E/J) \phi(x)$.   This form suggests a Wannier solution analogous to the above case, but with a two-site form factor to account for the rapidly oscillating factor $(-1)^x$. We propose the ansatz solution,
\be \phi_{ks}(x) \sim (1+A(-1)^{x+s})e^{-\pi\epsilon(x-x_{ks})^2} ,\label{psi_lat} \ee
where we have defined $x_{ks} = (k-s\pi)/(2\pi\epsilon)$ and $s = 0,1$ is a sub-band index.   Thus for each momentum $k$, there are two possible wave functions, which are spatially separated (due to the shift in the center of the oscillator) and also have their main weight on either the even or odd sites of the lattice.   Choosing the value $A = \sqrt{2}-1+\pi\epsilon(\sqrt{2}-2)/2$
 solves Harper's equation to order $O(\epsilon)$ and higher order terms can be added to $A$ to satisfy the equation to still higher order.

If we interpret the sub-band index $s$ as a quantum number representing a new degree of freedom, the low-energy bands in the lattice at $n_\phi = 1/2 + \epsilon$ are equivalent to the energy bands of a two-species system at $n_\phi = \epsilon$, which is the  continuum (Landau level) limit. Thus at
$n_\phi = 1/2 + \epsilon$ we can understand $\epsilon$ as the effective flux density giving rise to an effective filling fraction defined as $\tilde \nu = n/\epsilon$. Similarly, for general $n_\phi = p/q + \epsilon$ (with $p$ and $q$ coprime), $q$ solutions  can be found and the system can be treated as a $q$-species model with effective flux density $\epsilon$~\cite{palmer06,palmer08}.  With increasing $q$, the bandwidth increases as $\sim e^{-C/(q^2 \epsilon)}$ whereas the band gap, while remaining proportional to $\epsilon$, decreases with increasing $q$.

In order to be in the FQH regime, it is necessary that the interaction energy be larger than the bandwidth (so that the interaction dominates over the kinetic energy).   In addition, we would like the interaction to be smaller than the band gap so that all of the physics occurs within the lowest Landau band; however this requirement may not be crucial~\cite{LLMixing}.   Finally, the temperature must be less than the energy gap of the FQH state, which is typically set by the interaction energy (although it could also be set by the band gap if that is smaller). States competing with quantum Hall liquids include Bose-Einstein condensates: these describe the physics at $n_\phi=1/2$, for example~\cite{Moller10,Powell11}.

Given these restrictions, and given that experimentally obtaining low temperatures will always be a challenge, it is clear that
the FQH effect is most likely to be observed in the regime of intermediate $\epsilon$ where, the band gap is not too small and the bandwidth is not too large. Indeed, it is perhaps optimal to work in a regime where bandwidth and band gap are comparable. One can simply look at the Hofstadter spectrum to see where these inequalities are best satisfied~\cite{hofstadter76}.  The most experimentally favorable case occurs for $n_{\phi}=\epsilon \ll 1$. Here, $\epsilon$ might be as large as $0.3$ before the bandwidth is on the order of the band gap, and the band gap may be as large as $J$. This particular case has been studied extensively previously~\cite{sorensen05,hafezi07}.

The case of $n_{\phi} = 1/2 + \epsilon$, which we focus on here, is also fairly favorable for the observation of FQH effect. The parameter $\epsilon$ can be as large as $0.1$ before the band gap is on the order of the bandwidth, and the band gap may be as large as about $0.3J$. While $n_{\phi}  = 1/3 + \epsilon$ is still experimentally plausible, the cases of $n_{\phi}  = p/q + \epsilon$ with $q > 3$ have extremely tiny band gaps and, hence,   seem less accessible. We note that despite the fact that these inequalities of energy scales are   harder to satisfy for $n_{\phi} = 1/2 + \epsilon$ than for $n_\phi=\epsilon$, the former   has richer physics associated with the new quantum number, the sub-band index introduced above.

\begin{figure}[t]
\begin{center}
\includegraphics[width=1\columnwidth]{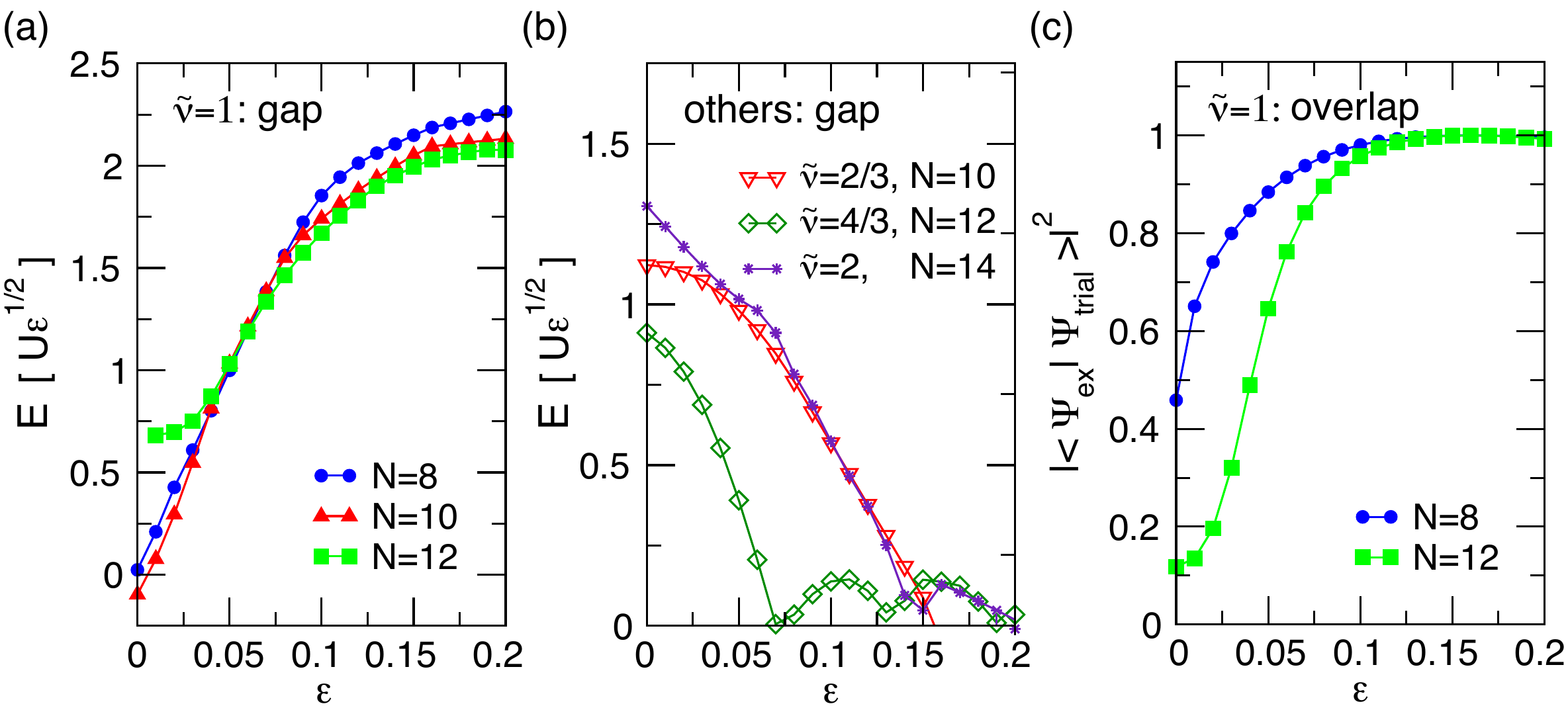}
\caption{(a) The FQH gap   $E$ at   effective filling fraction $\tilde \nu = 1$ as a function of $\epsilon$, where the flux density is $n_\phi = 1/2 + \epsilon$.   Data are shown for $N=8, 10$ bosons.  Increasing $\epsilon$ increases the strength of the pairing terms of the Hamiltonian and stabilizes this state. (b) The FQH gap for $\tilde \nu$ =  2/3 ($N = 10$), 4/3 ($N = 12$), and 2 ($N = 14$) as a function of $\epsilon$.   Increasing $\epsilon$ decreases the FQH gap and destabilizes the corresponding states. (c) Overlap between the exact ground state of the system at   effective filling fraction $\tilde \nu = 1$ and the trial wave function Eq.~(\ref{double pfaffian}) vs. $\epsilon$. The overlap for $\epsilon > 0.1$ exceeds $95\; \%$.}
\label{gap}
\end{center}
\end{figure}

We now turn to consider the effect of the interaction term in the Hamiltonian Eq.~(\ref{ham}).   Using any basis of (single-particle) states $\psi_a(x,y)$ with corresponding creation and annihilation operators $\hat \psi^\dagger_a$ and $\hat \psi_a$, the interaction may be decomposed as,
\be {\hat U} = \sum_{abcd} U_{abcd} \,\, \hat \psi_a^\dagger \hat \psi_b^\dagger \hat \psi_c^{\phantom{\dagger}} \hat \psi_d^{\phantom{\dagger}},\ee
where
\be  U_{abcd} = U  \sum_{x,y} \psi^*_{a}(x,y) \psi^*_{b}(x,y) \psi_{c}(x,y)\psi_{d}(x,y).\label{pseudopotential}\ee
For $n_\phi = \epsilon\ll1$, we use the basis for the lowest band, i.e., $\psi_k(x,y) = \phi_k(x)e^{iky}$ with $\phi_k(x)$ the Gaussian form as described above (properly normalized).   In this limit we may convert the sums into integrals, then, projected to the lowest energy band, we obtain $U_{k_1 k_2 k_3 k_4} = \sqrt{\epsilon}\, U e^{-\sum_{i<j=1}^4(k_i-k_j)^2/(16\pi\epsilon)} \delta_{k_1+k_2-k_3-k_4}$,   where the function $\delta_p$ is defined to be unity if the argument $p$ is an integer multiple of $2 \pi$ and is zero otherwise. The Gaussian factor enforces $k_i \approx k_j$ so that total momentum $k_1 + k_2 - k_3 - k_4$ must be zero, not just $0\mod2 \pi$. This derived form of the interaction is precisely what we expect for continuum bosons in the lowest Landau band with short-range contact interaction~\cite{Chakraborty}.

For $n_\phi = 1/2 + \epsilon$, we can use the basis $\psi_{ks}(x,y) = \phi_{ks}(x)e^{iky}$ with $\phi_{ks}(x)$ given by Eq.~(\ref{psi_lat}). Projecting to the lowest Landau level, we correspondingly find,
 \be \label{pseudopotential2} U_{k_1  s_1 k_2  s_2  k_3  s_3 k_4  s_4} =~~~~~~~~~~~~~~~~~~~~~~~~~~~~~~~~~~~~~~~~~~~~ \\\nn\sqrt{\epsilon}U  G_{s_1 s_2 s_3 s_4} e^{-\sum_{i<j=1}^4\frac{(k_i-k_j-\pi(s_i - s_j))^2}{16\pi\epsilon}}\delta_{k_1+k_2-k_3-k_4},
 \ee
where the matrix $G$ results from summing over the discrete form factors in the expressions for the corresponding wave functions.
  Note that the functional form of the interaction Eq.~(\ref{pseudopotential2}) is identical to contact interactions for a continuum Landau level up to the band index dependent matrix $G$ out front once we redefine the momentum as   $\tilde k =  k - \pi s$.  In terms of these new variables, the Gaussian enforces $\tilde k_i \approx \tilde k_j$ which now allows $k_1 + k_2 - k_3 - k_4 = \pm 2 \pi$ (allowed by $
\delta$) if  $s_1 + s_2 - s_3 - s_4 = \pm  2$.

Given the precise resemblance to a two-species continuum model, we may employ exact diagonalization, which is well established as a numerical technique for the study of interacting particles in   continuum Landau levels~\cite{Chakraborty}. We chose to formulate the effective problem obtained from our preceding analysis in a finite spherical geometry \cite{haldane83,haldane85}, which eliminates edge effects and thus allows direct access to the physics of the bulk. This problem is defined entirely within the lowest energy band by the interaction, Eq.~(\ref{pseudopotential2}), as the kinetic energy is assumed to have small bandwidth compared to the interaction scale $U$.

Following~\cite{palmer06} we switch to a new basis   $ \tilde \psi_{k\tilde s} = (\psi_{k0} + \tilde s \, i \psi_{k1})/\sqrt{2}$, where $\tilde s = \pm 1$ (or `up'  ($\uparrow$) and  `down' ($\downarrow$) in the current text).  We refer to this new form of the sub-band index $\tilde s$ as {\it pseudospin}. Using this basis, the nonzero elements of the transformed matrix $\tilde G$ become $ \tilde G_{\uparrow\uparrow\uparrow\uparrow} =  \tilde G_{\downarrow\downarrow\downarrow\downarrow} = \tilde G_{\uparrow\downarrow\uparrow\downarrow} = \tilde G_{\downarrow\uparrow\downarrow\uparrow} = \tilde G_{\downarrow\uparrow\uparrow\downarrow} = \tilde G_{\uparrow\downarrow\downarrow\uparrow} = 1$. In addition, we also find two extra ``pairing" terms: $\tilde G_{\downarrow\downarrow\uparrow\uparrow} = \tilde G_{\uparrow\uparrow\downarrow\downarrow} = \pi\epsilon$. These terms correspond to $k_1+k_2 = k_3+k_4+2\pi$, which resemble \emph{umklapp} scattering processes, and do not conserve pseudospin $(\tilde{s}_1+\tilde{s}_2 \neq \tilde{s}_3+\tilde{s}_4)$. These terms indicate that a pair of pseudospin ups (downs) can be annihilated while a pair of pseudospin downs (ups) are created thus suggesting that a BCS-type pairing~\cite{bcs,generalPairing} could occur between particles with the same pseudospin.  While these  pairing terms vanish in the limit of $\epsilon \rightarrow 0$, as mentioned above, the experimentally relevant regime is likely to be at finite $\epsilon$   where these terms will be important.

The novel twist for lattice bosons near $n_\phi=1/2$ is the emergence of an umklapp scattering term between the two emergent species. Consequently, the total pseudospin is not conserved, and we need to take into account the full Hilbert space containing all possible distributions of particles into the two sub-bands. We examine the spectrum of the Hamiltonian Eq.~(\ref{pseudopotential2}) for the occurrence of incompressible ground states that are characterized by translational invariance (angular momentum $L=0$ on the sphere) and a finite FQH gap. Our search yields four candidates at effective filling fractions $\tilde{\nu} =  n/\epsilon = 2/3, 1, 4/3$ and $2$ where $n_\phi = 1/2 + \epsilon$~\cite{shifts}.

For the effective filling fraction $\tilde{\nu} = 1$, we find that the energy gap between the ground state and the first excited state rapidly opens up as one increases $\epsilon$ [see Fig.~\ref{gap}(a)].    This indicates that the umklapp pairing terms, $\tilde G_{\downarrow\downarrow\uparrow\uparrow}$  and $\tilde G_{\uparrow\uparrow\downarrow\downarrow}$, which are the only terms in Eq.~(\ref{pseudopotential2}) that change with $\epsilon$, are responsible for producing an incompressible state at this filling factor. On the contrary, the energy gaps at effective filling fractions $\tilde{\nu} = 4/3$ and $\tilde{\nu} = 2$ close as $\epsilon$ increases, indicating that the pairing terms {\it destabilize} the corresponding incompressible states [see Fig.~\ref{gap}(b)].  At $\tilde{\nu} = 2/3$, as pointed out in~\cite{palmer06}, the 221 state is an exact ground state, and this remains true even in the presence of the  pairing terms.   However, as $\epsilon$ increases, this gap also closes, as excited states are sensitive to   the  pairing terms [see Fig.~\ref{gap}(b)].

Interestingly, the energy gaps of the $\tilde\nu=2/3$ and $\tilde\nu=2$ states have a very similar magnitude and dependency on $\epsilon$, which may be unexpected in the current formalism.  This, however, is a natural conclusion in the composite fermion (CF) theory for the lattice, which explains both these states by the same energy gap in the CF spectrum~\cite{moller09}. Furthermore, our expansion of the effective model to linear order in $\epsilon$ predicts that the gap of these states closes near the value, $\epsilon=1/6$, predicted by CF theory~\cite{moller09}. The case of $\nu=4/3$ is clearly very different. Here we find that the ground state at small $\epsilon$ has significant overlap of $\gtrsim 0.7$ with the non-Abelian spin singlet (NASS) state~\cite{NASS} for $N=12$ particles. However, this state is very fragile and we cannot ascertain that it persists in the thermodynamic limit.

As was mentioned above, the presence of the umklapp pairing terms in the Hamiltonian, $\tilde G_{\downarrow\downarrow\uparrow\uparrow}$  and $\tilde G_{\uparrow\uparrow\downarrow\downarrow}$, suggest a BCS-type pairing between particles of the same pseudospin. To further investigate the effect of these terms on the nature of the ground state at $\tilde{\nu} =1$ we propose an explicit trial wave function.  We use the common conventions for studying the FQH effect: we adopt the symmetric gauge and write coordinates of particles on the plane in dimensionless complex form $z = (x + i y )/l_0$. Our trial wave function is given in the coordinates of $N_\uparrow$ bosons of `up' type and $N_\downarrow$ bosons of `down' type (with both $N_\uparrow$ and $N_\downarrow$ assumed to be even):
\be \nn
&  & \Psi(\{z^\uparrow_i\},\{z^\downarrow_j\}) = {\rm Pf}\left(\frac{1}{z^\uparrow_i-z^\uparrow_j}\right){\rm Pf}\left(\frac{1}{z^\downarrow_i-z^\downarrow_j}\right)\\\nn&\times& \prod_{i<j=1}^{N_\uparrow}(z^\uparrow_i - z^\uparrow_j)\prod_{i<j=1}^{N_\downarrow}(z^\downarrow_i - z^\downarrow_j)\prod_{i=1}^{N_\uparrow}\prod_{j=1}^{N_\downarrow}(z^\uparrow_i - z^\downarrow_j)\\&\times&e^{-\sum_{i=1}^{N_\uparrow} |z_i^\uparrow|^2/4 -\sum_{i=1}^{N_\downarrow} |z_i^\downarrow|^2/4}.\label{double pfaffian}
\ee
As written, the up and down bosons are assumed distinguishable. Expanding this wave function in the original bosonic basis, one obtains an expression that is fully symmetric in all coordinates.

Here, the two Pfaffian factors (Pf) are antisymmetrized sums over pairs of particles with the same pseudospin ${\rm Pf}(\frac{1}{z_i-z_j}) = \mathcal{A}(\frac{1}{z_1-z_2}\frac{1}{z_3-z_4}\cdots)$ with ${\cal A}$ the antisymmetrizing operator.  This Pfaffian form is precisely the real space form of a BCS pairing wave function, which indicates that particles with the same pseudospin form pairs. Without the Jastrow factors this type of pairing is analogous to a $^3$He-A phase with a $d$-vector in the $x,y$-plane~\cite{bcs}.  As in the case of other paired Hall states, the topological properties are only trivially altered by restoring the Jastrow factors~\cite{bcs}.

Note that the wave function, Eq.~(\ref{double pfaffian}), is the exact ground state of $\mathcal{H}_{2-3}$, the sum of a three-body delta-function interaction for particles with the same pseudospin and a two-body delta-function interaction between particles of opposite pseudospin.

To check the validity of this trial wave function (obtained as the ground state of $\mathcal{H}_{2-3}$ on the sphere), we calculate its overlap with projection of the ground state of Eq.~(\ref{pseudopotential2}) onto the sector with $N_\uparrow=N_\downarrow$. As $\epsilon$ is increased and the gap opens up, we find increasing overlap between our trial state and the exact ground state [see Fig.~\ref{gap}(c)]. The overlap for $\epsilon=0.1$ is above $95\; \%$ for $N=12$ particles, which is an excellent indicator of the validity of our proposed wave function. Note that, although outside the regime of validity for our model, at $\epsilon\simeq 0.16$, Eq.~(\ref{double pfaffian}) is nearly an exact ground state of the two-body interaction, Eq.~ (\ref{pseudopotential2}), to an accuracy of about $10^{-5}$. We also find that the inverse  $1/z$ dependence of the paired wave function in Eq.~(\ref{double pfaffian}) is optimal, as introducing variational parameters to change its shape~\cite{generalPairing} does not increase the overlap significantly.

We have also studied the quasihole spectrum of Eq.~(\ref{double pfaffian}) in the presence of additional flux.  For the model $\mathcal{H}_{2-3}$ Hamiltonian, the quasihole spectrum is precisely that of two decoupled Moore-Read layers --- the quasiholes of each layer corresponding to the so-called half-quantum vortices of $^3$He-A.  However, for our Hamiltonian of interest, Eq.~(\ref{pseudopotential2}), the umklapp pairing terms lock the direction of the $d$-vector, thus requiring that the quasiholes pair between layers, confining the half quantum vortices and leaving the system with effectively Abelian excitations.  To establish with clarity that this is the correct physics we have been able to predict the entire low-energy quasihole spectrum of the $^3$He-A model, using a generalization of the approach introduced in~\cite{read96}, which precisely matches the low-lying spectrum of the microscopic Hamiltonian, Eq.~(\ref{pseudopotential2}), for every case we could numerically access.  These results will be presented elsewhere. 

Signatures for our proposed state can be derived from a range of experimental probes for the detection of quantum Hall states in cold gases, such as measurements of groundstate incompressibility~\cite{cooper05}, noise correlations~\cite{altman04, palmer08}, and possibly a direct measurement of quasihole statistics~\cite{paredes01}.

The methods described here can be generalized to flux density $n_{\phi} = p/q + \epsilon$ although, as discussed above, FQH effect with larger $q$ is likely to be harder to realize in experiments.  In this case there would be a sub-band index $s = 0,1 \ldots (q-1)$ and the umklapp terms of the interaction would allow nonconservation of this sub-band index via  $s_1+s_2 - s_3- s_4 =0 \mod q$, which could lead to new pairing terms and possibly new physics.

To summarize, we have shown that  anomalous pairing (umklapp) interaction terms are crucial to the physics of FQH effect for interacting bosons on a lattice at  flux density $n_\phi = 1/2 + \epsilon$.  We find that the pairing terms greatly modify the ground state at various effective filling fractions $\tilde \nu = n/\epsilon$.  At $\tilde \nu=1$, we demonstrate that these terms stabilize a new paired FQH state, which is effectively two coupled copies of the Moore-Read Pfaffian state.  At $\tilde \nu=2/3, 4/3$ and $2$, we find that the incompressible states are destabilized by the  pairing terms.
\begin{acknowledgements}
Discussions with S. Adam, E. Ardonne, N.~R.~Cooper, M. Hafezi, L. Mathey, R. Palmer, M. Peterson, and especially E. Tiesinga are gratefully acknowledged. The authors acknowledge the hospitality of Nordita and the Aspen Center for Physics and support from NIST/NRC (L.H.), Trinity Hall Cambridge, the Newton Trust, and the Leverhulme Trust under Grant ECF-2011-565 (G.M.) and EPSRC Grant EP/I032487/1 (S.H.S.).
\end{acknowledgements}


\begin{references}
\bibitem{LLMixing} It was found in Ref.~\cite{hafezi07} that certain FQH states of bosons may persist even when the interaction scale is much larger than the band gap.  Similarly in most semiconductor samples many FQH states persist even when the interaction scale is substantially larger than  the (cyclotron) band gap.
\bibitem{shifts} FQH states have an additional quantum number $\sigma$, known as shift, such that $N_\phi=\tilde\nu^{-1}N-\sigma$ in finite spherical systems. Our states have $(\tilde\nu,\sigma)=(2/3,3), (1,2), (4/3,2)$, and $(2,2)$, respectively.
\end{references}
\end{document}